\documentclass{PoS}
\usepackage{booktabs}
\usepackage{amsmath}
\usepackage{multirow}
\usepackage{textcomp}
\usepackage{graphicx}
\usepackage{subcaption}
\usepackage{selinput}
\SelectInputMappings{
   aacute={á},
   ntilde={ñ}
}

\title{$D$ meson Semileptonic Decay Form Factors at $q^2 = 0$}

\ShortTitle{$D$ meson Semileptonic Decay Form Factors at $q^2 = 0$}

\author{
\speaker{Ruizi Li}\nolinebreak$^a\thanks{Current address: Department of Physics and Astronomy, Michigan State University,
East Lansing, MI 48824, USA}$,
A.~Bazavov$^{b,c}$,
C.W.~Bernard$^d$,
C.~DeTar$^e$,
D.~Du$^f$,
A.X.~El-Khadra$^{g,l}$,
E.~G\'amiz$^h$,
S.~Gottlieb$^a$,
U.M.~Heller$^i$,
J.~Komijani$^j$,
A.S.~Kronfeld$^{k,l}$,
J.~Laiho$^f$,
P.B.~Mackenzie$^l$,
E.T.~Neil$^{m,n}$,
T.~Primer$^o$,
J.N.~Simone$^l$,
R.L.~Sugar$^p$,
D.~Toussaint$^o$,
R.S.~Van~de~Water$^l$, and
R.~Zhou$^l$
\\
\llap{$^a$} Department of Physics, Indiana University, Bloomington, IN 47405, USA\\
\llap{$^b$} Department of Computational Mathematics, Science and Engineering, Michigan State University, East Lansing, MI 48824, USA\\
\llap{$^c$} Department of Physics and Astronomy, Michigan State University, East Lansing, MI 48824, USA\\
\llap{$^d$} Department of Physics, Washington University, St. Louis, MO 63130, USA\\
\llap{$^e$} Department of Physics and Astronomy, University of Utah, Salt Lake City, UT 84112, USA\\
\llap{$^f$} Department of Physics, Syracuse University, Syracuse, NY 13244, USA\\
\llap{$^g$} Department of Physics, University of Illinois, Urbana,  IL 61801, USA\\
\llap{$^h$} CAFPE and Departamento de F\'{\i}sica Te\'orica y del Cosmos, Universidad de Granada, E-18071 Granada, Spain\\
\llap{$^i$} American Physical Society, One Research Road, Ridge, NY 11961, USA\\
\llap{$^j$} School of Physics and Astronomy, University of Glasgow, Glasgow G12 8QQ, UK\\
\llap{$^k$} Institute for Advanced Study, Technische Universit{\"a}t M{\"u}nchen, 85748 Garching, Germany\\
\llap{$^l$} Fermi National Accelerator Laboratory, Batavia, IL 60510 USA\\
\llap{$^m$} Department of Physics, University of Colorado, Boulder, CO 80309, USA\\
\llap{$^n$} RIKEN-BNL Research Center, Brookhaven National Laboratory, Upton, NY 11973, USA\\
\llap{$^o$} Physics Department, University of Arizona, Tucson, AZ 85721, USA\\
\llap{$^p$} Department of Physics, University of California, Santa Barbara, CA 93106, USA

\vspace{2mm}
{\large\bf Fermilab Lattice and MILC Collaborations}
\vspace{3mm}

E-mail:
\email{ruizli@umail.iu.edu}
}

\abstract{
We discuss preliminary results for the vector form factors $f_+^{\{\pi,K\}}$ at zero-momentum transfer for the decays $D\to\pi\ell\nu$ and $D\to K \ell\nu$ 
using MILC's $N_f = 2+1+1$ HISQ ensembles at four lattice spacings, 
$a \approx 0.042, 0.06, 0.09$, and 0.12~fm, and various HISQ quark masses down to the (degenerate) physical light quark mass. 
We use the kinematic constraint $f_+(q^2)= f_0(q^2)$ at $q^2 = 0$ to determine the vector form factor from our study of the scalar current, which yields $f_0(0)$. 
Results are extrapolated to the continuum physical point in the framework of hard pion/kaon SU(3) heavy-meson-staggered $\chi$PT and Symanzik effective theory. 
Our calculation improves upon the precision achieved in existing lattice-QCD calculations of the vector form factors at $q^2=0$. 
We show the values of the CKM matrix elements $|V_{cs}|$ and $|V_{cd}|$ that we would obtain using our preliminary results for the form factors together with recent experimental results, 
and discuss the implications of these values for the second row CKM unitarity. }

\FullConference{The 36th Annual International Symposium on Lattice Field Theory - LATTICE2018\\
		22-28 July, 2018\\
		Michigan State University, East Lansing, Michigan, USA.}

\begin{document}

\section{Motivation}

In the Standard Model (SM), the Cabibbo-Kobayashi-Maskawa (CKM) matrix must be unitary. 
This property has been subjected to more and more stringent tests in order to probe for evidence of physics beyond the SM.
Lattice QCD provides crucial theoretical input, 
e.g., leptonic decay constants and semileptonic decay form factors. 
If the experimental measurements for two processes that depend on the 
same CKM matrix element imply different values for that element, that would
provide evidence for new physics. 
In this way, SM predictions constrain contributions from new physics. 
Lattice-QCD calculations with improved precision of the $D$ meson semileptonic decay vector form factors, $f_+^\pi$ and $f_+^K$, 
are important for a more precise determination of the second row 
CKM matrix elements $|V_{cd}|$ and $|V_{cs}|$, 
and a more stringent test of second row CKM unitarity, 
since current determinations are limited by the form factors errors. 
In particular, as shown in Table~\ref{table:1}, 
the error in recent lattice-QCD calculations of $f_+^{\{\pi,K\}}$ contributes roughly 3 to 4 times that of the experimental error of the decay rate. 
\begin{table}[htbp]%[htdp]
\begin{center} 
\begin{tabular}{c c c}
\hline\hline
 & $\mid V_{cd} \mid$ & $\mid V_{cs} \mid$ \\
\hline
Semileptonic & 0.2140(93)(29) & 0.975(25)(7) \\
\hline
Leptonic & 0.2164(14)(49) & 1.008(5)(16) \\
\hline\hline
\end{tabular}
\end{center}
\caption{The CKM matrix second row elements $|V_{cd}|$ and $|V_{cs}|$ in the FLAG report~\cite{FLAG}, 
determined from semileptonic and leptonic decay processes. 
The first and second error in each entry are from lattice-QCD calculations and experiments respectively. }
\label{table:1}
\end{table}

\section{Lattice-QCD calculations of $f_+^\pi$ and $f_+^K$  at zero-momentum transfer}

Form factors for the hadronic $D$ meson semileptonic decay are defined via the heavy-light vector current matrix element: 
\begin{eqnarray}
\langle P(p) | \bar{l} \gamma_\mu c | D(p') \rangle = f^P_+(q^2) [ (p' + p)_\mu - \frac{m_D^2 - m_P^2}{q^2} q_\mu] + f^P_0(q^2) \frac{m_D^2 - m_P^2}{q^2} q_\mu ,
\end{eqnarray}
where $q = p' - p$ is the hadron momentum transfer, 
$l$ is the active light quark in the current, 
and $P$ is the daughter-meson, $\pi$ or $K$. 
At zero-momentum transfer, 
and using the kinematic constraint $f_+(0) = f_0(0)$, 
we just need to calculate the scalar form factor $f_0 (0)$ with the scalar current insertion, 
\begin{eqnarray}
f^P_0(q^2) = \frac{m_c - m_l}{m_D^2 - m_P^2} \langle P(p) | \bar{l}c | D(p') \rangle . 
\end{eqnarray}
With lattice fermions, this scalar density (defined to include the explicit quark-mass factor) is absolutely normalized.

\subsection{Lattice setup} 
Our simulations are done on dynamical four-flavor highly-improved staggered quark (HISQ) ensembles~\cite{HISQ} generated by the MILC collaboration, 
with 
degenerate up and down quarks, and both strange and charm
quarks tuned close to their physical masses. 
Lattice spacings vary from $a \approx 0.042$~fm to $a \approx 0.12$~fm, 
and up/down quark masses are as low as, and in some cases even lower than, their physical value. 
For the up and down quarks, the valence and sea quark masses are identical; 
however, for the strange and charm better tuned masses than those in the sea have been used in the valence sector. 
Three meson source-sink temporal separations in 
three-point correlators are 
used on most ensembles to further improve fits. 
Table~\ref{table:2} lists all the 
HISQ ensembles used in this calculation. 
Twisted boundary conditions are imposed to achieve the daughter-quark momentum, $\vec{p} = p_i (1,1,1)$, 
with an equal component along each spatial direction so $q^2 = (p' - p)^2 = 0$. 
We use a wall quark source and point sink for an enhanced overlap with the ground state, 
and further fix the $D$ meson to be at rest.  
\begin{table}[htbp]%[htdp]
\begin{center}
\begin{tabular}{|c|c|c|c|c|c|}
\hline
Volume &	$a$ (fm)	&	$m_l / m_s$	&	$N_\text{conf}\times N_\text{src}$	&	$m'_c/m_c$	&	$T_\text{ext}$	\\
\hline
$63^3 \times 192$	&	0.042	&	0.2	&	$431 \times 12$	&	1.00	&	40	\\
\hline
$96^3 \times 192$	&	0.06	&	Phys.	&	$866 \times 6$	&	1.01	&	31, 39, 40	\\
\hline
$48^3 \times 144$	&	0.06	&	0.2	&	$942 \times 8$	&1.11	&	34, 41, 48	\\
\hline
$64^3 \times 96$	&	0.09	&	Phys.	&	$905 \times 8$	&	1.00	&	23, 27, 32	\\
\hline
$48^3 \times 96$	&	0.09	&	0.1	&	$840 \times 8$	&	1.02	&	23, 27, 32	\\
\hline
$32^3 \times 96$	&	0.09	&	0.2	&	$645 \times 4$	&	1.04	&	23, 27, 32	\\
\hline
$48^3 \times 64$	&	0.12	&	Phys.	&	$942 \times 4$	&	0.98	&	15, 18, 20	\\
\hline
$32^3 \times 64$	&	0.12	&	0.1	&	$992 \times 4$	&	1.02	&	15, 18, 20	\\
\hline
$24^3 \times 64$	&	0.12	&	0.2	&	$1050 \times 4$	&	1.00	&	15, 18, 20	\\
\hline
$24^3 \times 64$	&	0.12	&	0.1	&	$1018 \times 8$	&	1.02	&	15, 18, 20	\\
\hline
$40^3 \times 64$	&	0.12	&	0.1	&	$1001 \times 8$	&	1.02	&	15, 18, 20	\\
\hline
\end{tabular}
\end{center}
\caption{HISQ ensembles used in this calculation. 
The fifth column is the ratio of the simulation charm-quark mass, $m'_c$, to the physical one. 
The last column gives the $D$ and daughter-meson temporal separations in three-point correlators. }
\label{table:2}
\end{table}

\subsection{Data and fits}
Meson propagators are in general correlated among configurations, 
thus we fix the block size on different ensembles to be 2, 3, or 4, 
to eliminate the impact of any autocorrelations on the fitted result. 
We simultaneously fit the D and daughter-meson 2-point functions, 
and the 3-point functions for all values of $T_\text{ext}$ using a Bayesian method. 
We use 2 or 3 excited states of both positive and negative parities in our fitting function to efficiently disentangle the excited-state contributions, 
with the exception of the zero-momentum pion propagator where no negative parity eigenstates exist. 
Fitting windows are aligned at the meson source end on all ensembles, with $t_\text{min} \approx 0.36$~fm and 0.45~fm for the daughter meson and $D$ meson respectively. 
We use a jackknife analysis to determine the statistical errors. 
For a more detailed discussion on the correlator fits, see Ref.~\cite{D2Kpi2016}. \\
\begin{figure}[htbp]
\centering
 \begin{subfigure}[b]{0.5\textwidth} 
 \centering
  \includegraphics[width=\textwidth]{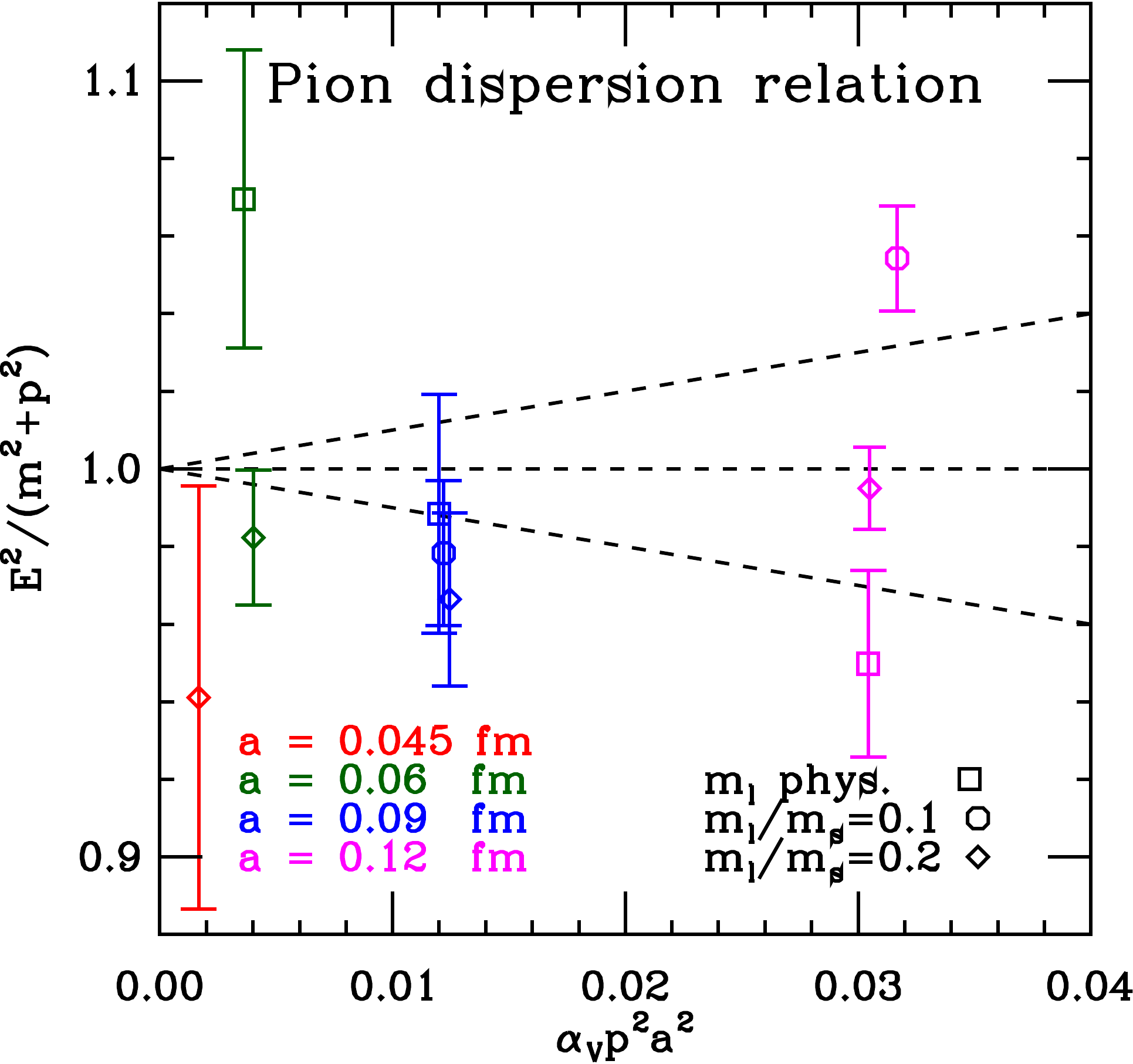}
  \label{a}
 \end{subfigure}
% \hspace{0.1cm}
 \begin{subfigure}[b]{0.5\textwidth} 
 \centering
  \includegraphics[width=\textwidth]{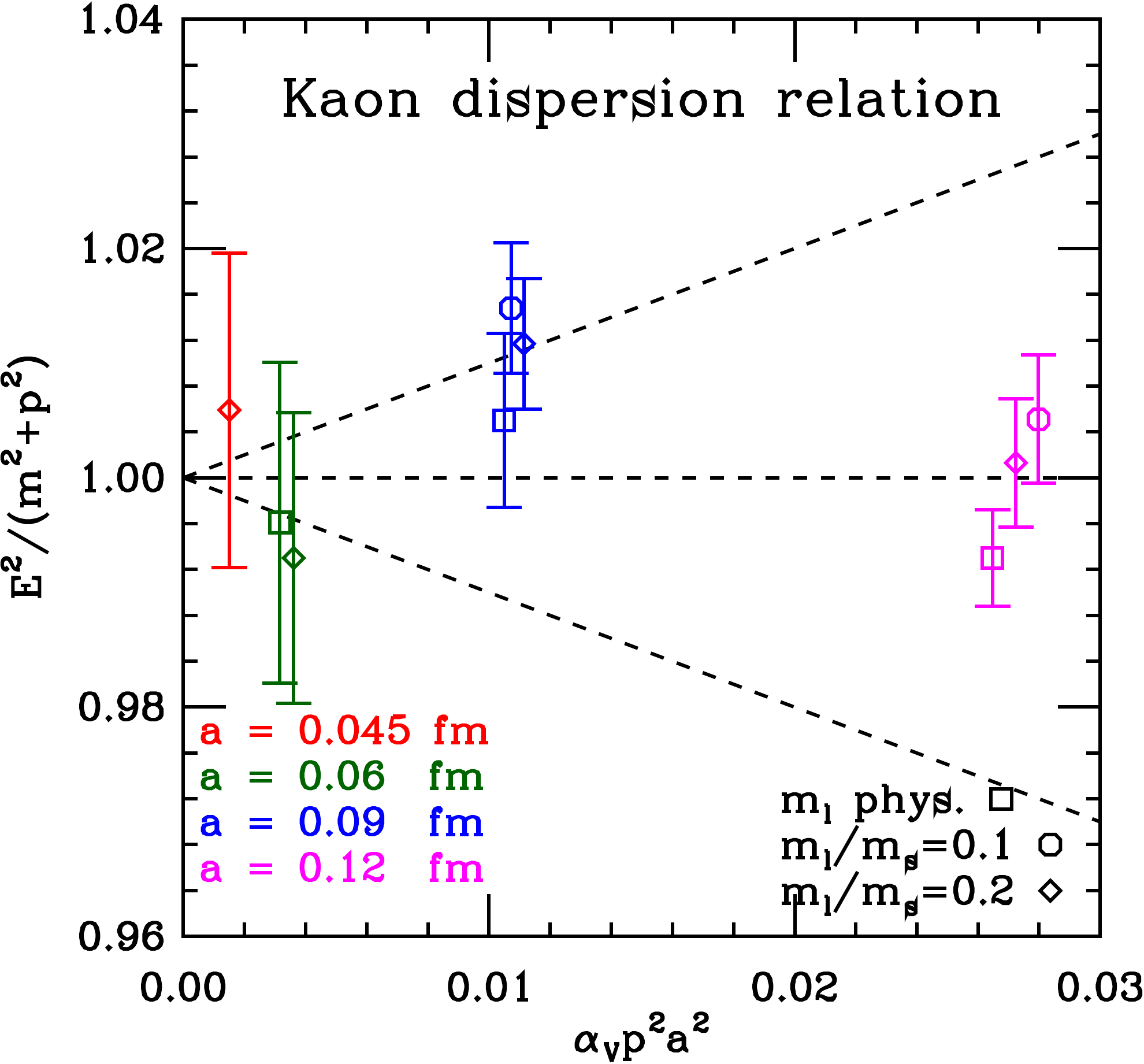}
  \label{b}
  \end{subfigure} 
\caption{Plots of pion (up) and kaon (down) dispersion relations. 
Different shapes and colors of data points represent different light quark masses and lattice spacings separately. }
\label{Fig:1}
\end{figure}
Figure~\ref{Fig:1} shows the pion and kaon energy dispersion relation on each ensemble. 
Dotted lines with the slope are to guide the eye on lattice discretization effects which scale as $\alpha_s p^2 a^2$. 
Deviations from the dispersion relation $E^2=\vec{p}^2+m^2$ are due to both the lattice discretization and the statistical error affected by the twist angle set up at an earlier stage. 
As shown in the plots, such deviations in the pion are more significant than those in the kaon.

\subsection{Chiral-continuum extrapolations}

We apply SU(3) hard-pion(kaon) heavy-meson rooted-staggered chiral perturbation theory (HMrS$\chi$PT)~\cite{Chi1,Chi2,Chi3,Chi4} for chiral-continuum extrapolations of the scalar form factor $f_0^{\{\pi, K\}}$ at $q^2 = 0$. 
The chiral formula includes the one-loop logarithmic term, 
as well as analytic terms in expansions of, 
e.g., valence and sea light quark masses, $q^2$, and the daughter meson energy,  
up to next-to-next leading order (NNLO), 
\begin{eqnarray}
f_0^P(0) = \frac{f_{p4s}C_0}{f_\pi} ( 1 + \delta f_{cl} + C_a \chi_a + C_l \chi_l + C_q \chi_q + C_s \chi_s + C_e \chi_e + ... ), 
\end{eqnarray}
where $\delta f_{cl}$ is the one-loop chiral logarithmic term, 
\begin{eqnarray}
\chi_a &=& (8\pi^2 f_{\pi}^2)^{-1} \bar{\Delta}(a), \label{eqn1} \\
\chi_l &=& (4\pi^2 f_{\pi}^2)^{-1} m_{u(d), s} \mu(a), \label{eqn2} \\
\chi_q &=& (8\pi^2 f_{\pi}^2)^{-1} q^2, \label{eqn3} \\
\chi_s &=& (8\pi^2 f_{\pi}^2)^{-1} (2 m_{u(d)} + m_s') \mu(a), \label{eqn4} \\
\chi_e &=& \sqrt{2} (4\pi f_{\pi})^{-1} E_P , \label{eqn5}
\end{eqnarray} 
$f_{p4s}$ is defined as the fictitious pseudoscalar decay constant with degenerate valence quarks of mass $m_{u(d)} = 0.4 m_s$ and physical sea-quark masses~\cite{Fp4s}, 
$\bar{\Delta}(a)$ is the average taste splitting, and $\mu(a)$ is the the low-energy constant that relates meson and quark masses in $\chi$PT. 
The ellipsis contains contributions from, 
e.g., light meson energy and momentum discretizations, 
the charm quark mass mistuning, and NNLO terms, 
that we use to check the systematic errors in our analysis. 

\begin{figure}[htbp]
\centering
 \begin{subfigure}[b]{0.5\textwidth} 
  \includegraphics[width=\textwidth]{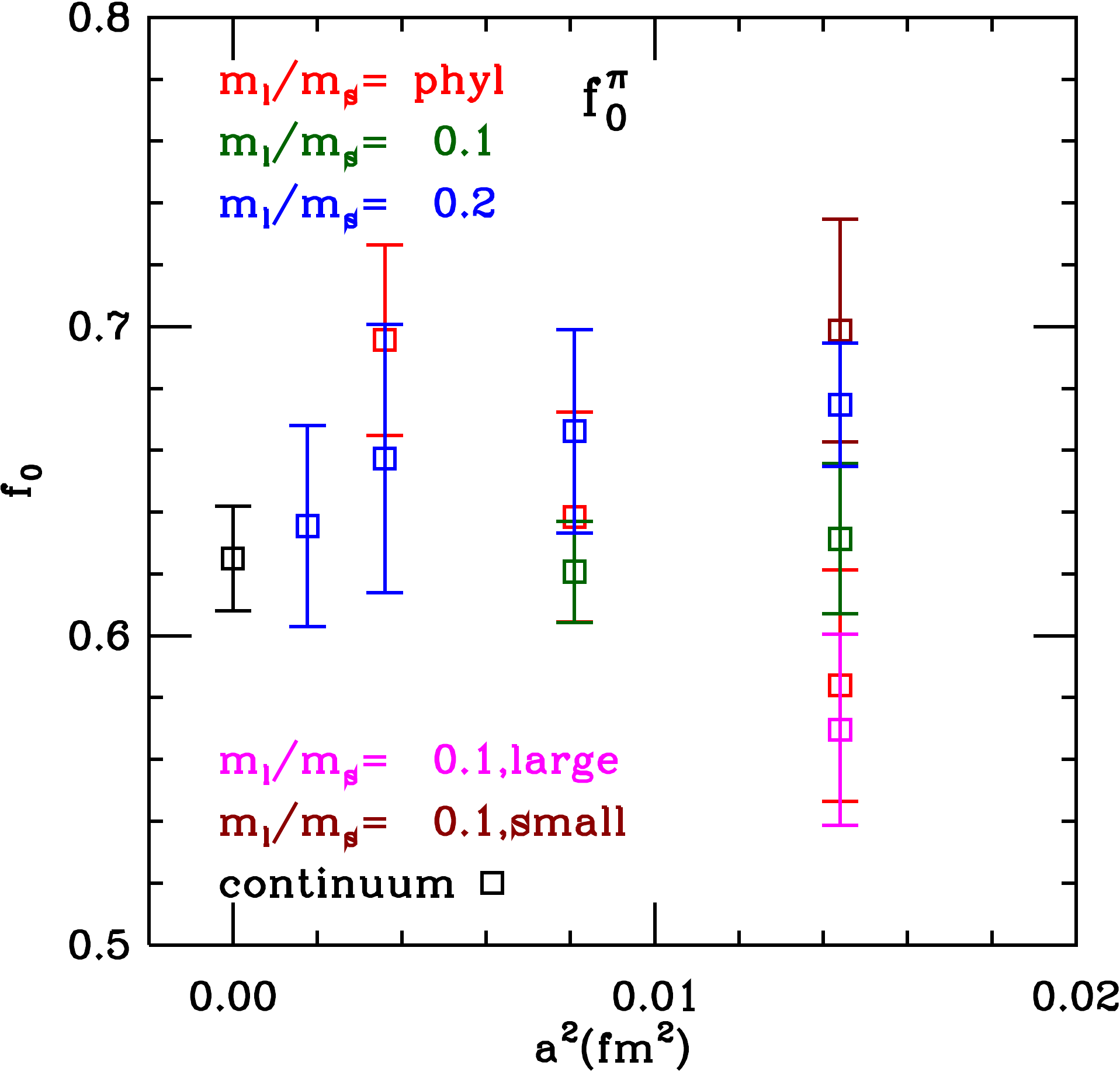}
  \label{a}
 \end{subfigure}
% \hspace{0.1cm}
 \begin{subfigure}[b]{0.5\textwidth} 
  \includegraphics[width=\textwidth]{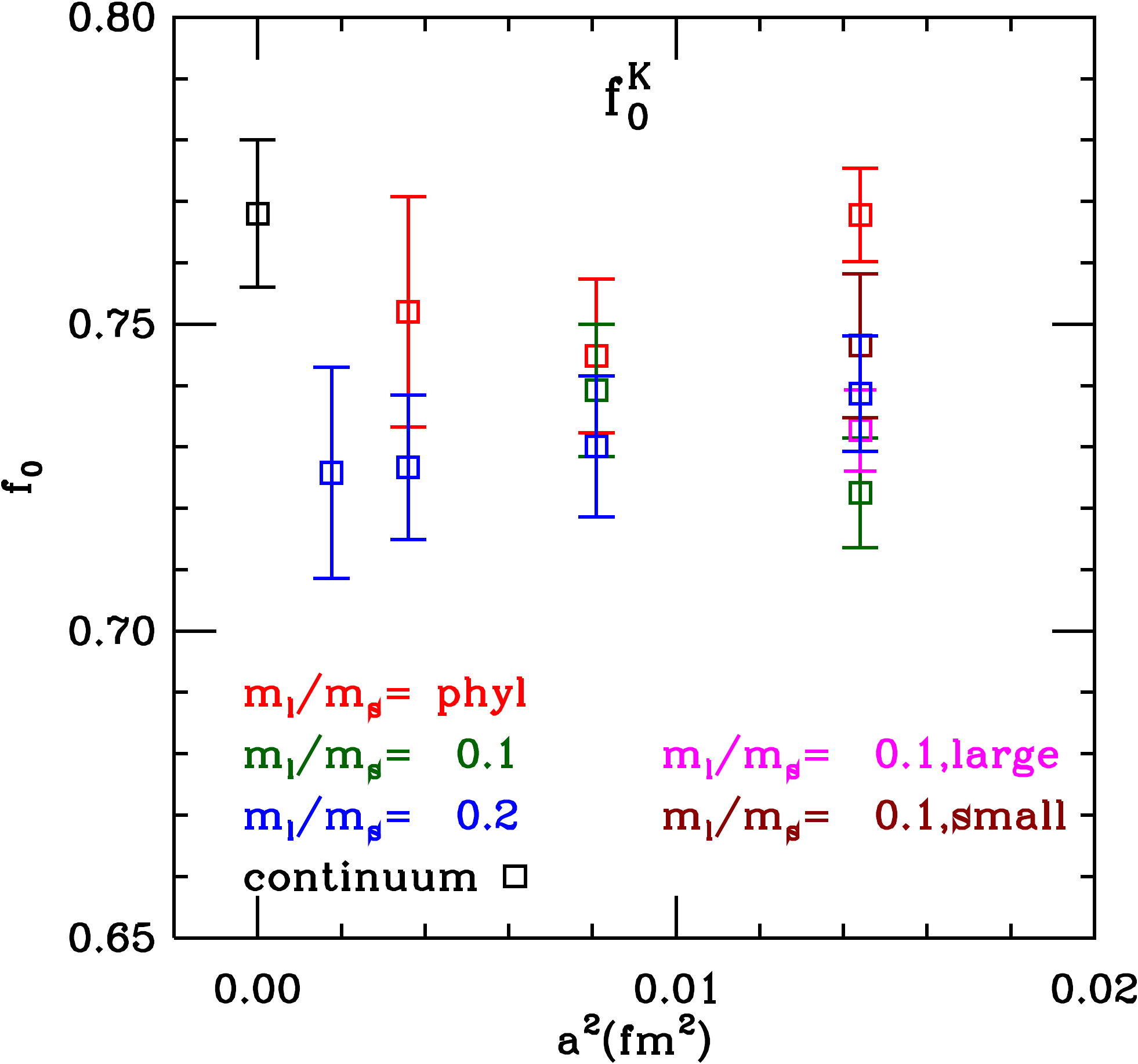}
  \label{b}
  \end{subfigure} 
\caption{Plots of fitted $f_0^{\pi}$ (up) and $f_0^K$ (down) on each ensemble and at the physical point in the continuum limit (points in black). 
The $x$-axis is the squared lattice spacing. 
Data in red, green, and blue are on ensembles with the light quark mass $m_l$ equals the physical value, 
$0.1m_s$, and $0.2 m_s$ respectively. 
Data in brown and pink are on coarse ensembles with a smaller- and larger-than-average spatial volume. }
\label{Fig:2}
\end{figure}

\begin{figure}[htbp]
\centering
 \begin{subfigure}[b]{0.5\textwidth} 
  \includegraphics[width=\textwidth]{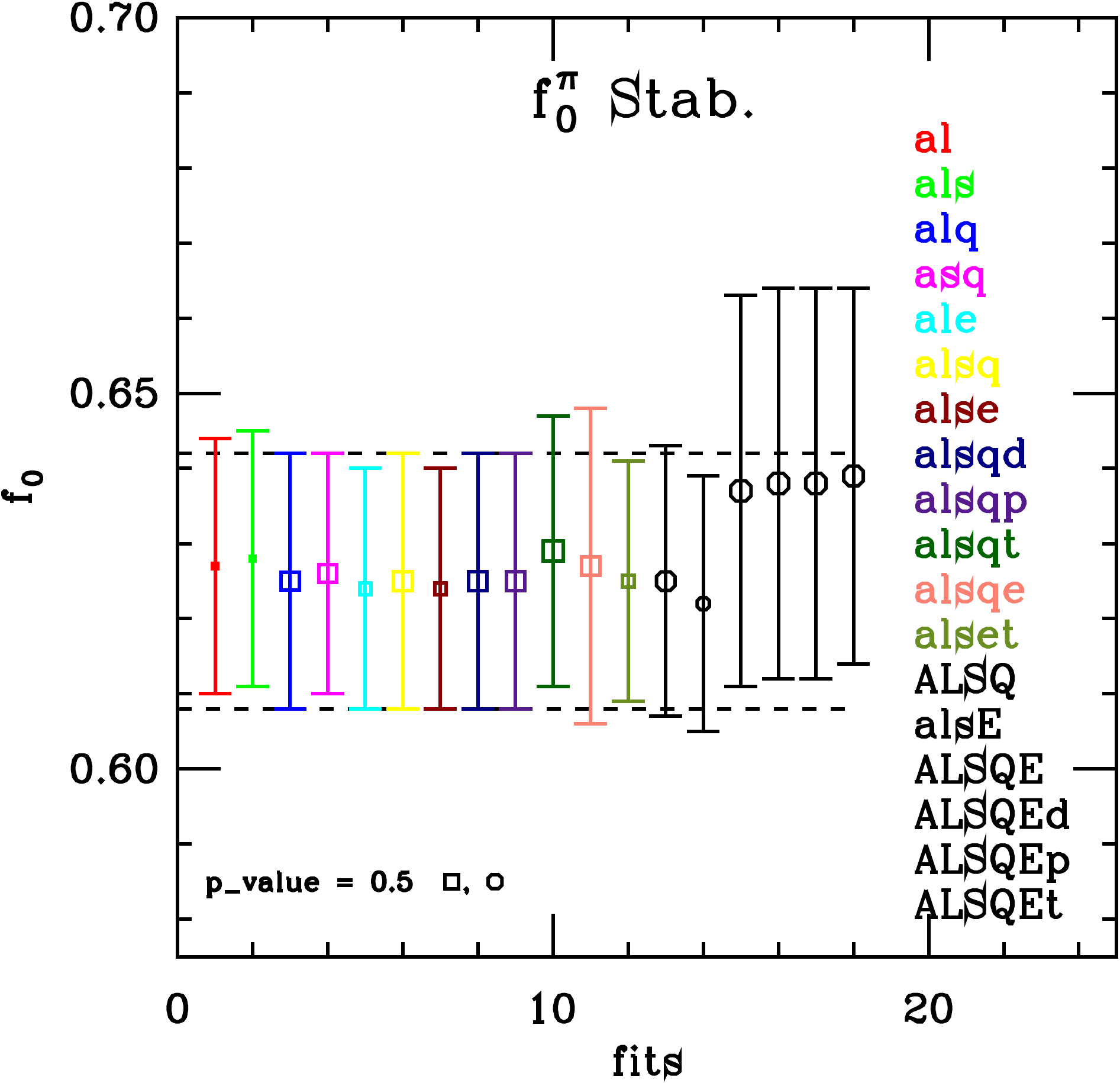}
  \label{a}
 \end{subfigure}
% \hspace{0.1cm}
 \begin{subfigure}[b]{0.5\textwidth} 
  \includegraphics[width=\textwidth]{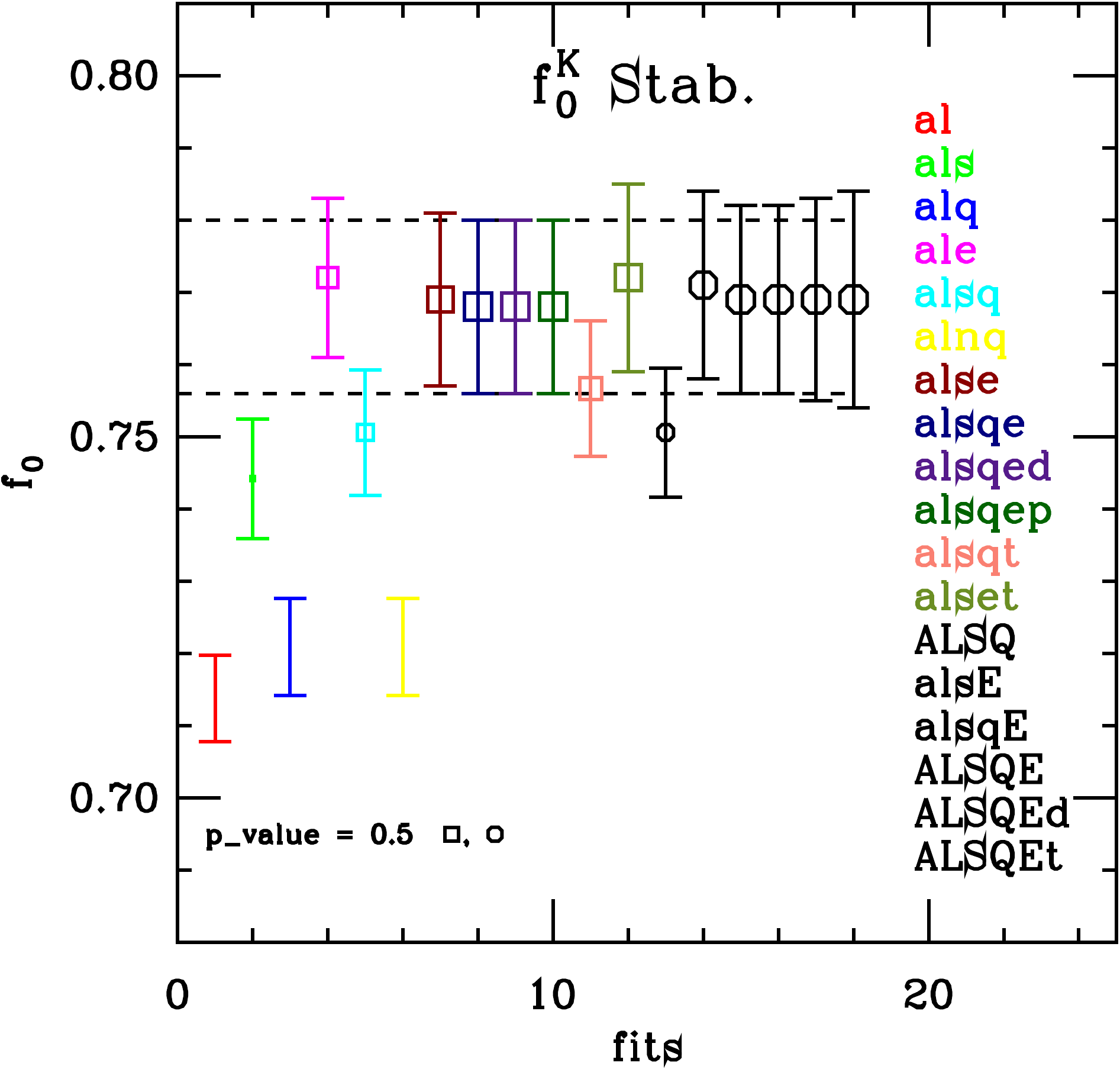}
  \label{b}
  \end{subfigure} 
\caption{Plots of the stability of the chiral-continuum fits of $f_0^{\pi}$ (up) and $f_0^K$ (down). 
Points from left to right corresponds to fitting terms, listed in right of each plot, from top to bottom. 
Besides listed in Eqn.~\ref{eqn1} -- \ref{eqn5}, 
the notation of the analytical term is: 
d -- daughter meson energy discretization; 
p -- daughter meson spatial momentum discretization; 
n -- valence and sea strange quark mass difference; 
t -- charm quark mass mistuning. 
Capital letters are for NNLO terms. 
The symbol size is proportional to the p-value of the fit. 
}
\label{Fig:3}
\end{figure}

Figure~\ref{Fig:2} shows the form factor $f_0^{\{\pi, K\}}(0)$ on each ensemble and at the physical point in the continuum limit. 
Together with our central fit, marked with dotted lines, 
Figure~\ref{Fig:3} shows the results of several alternate fits, up to NNLO, 
that we use to check the stability of our fitting procedure. 
For both form factors, fits that have reasonable $\chi^2/dof$ give results that agree with the central-fit results within errors. 
Central fits here are taken as $f_0^\pi = 0.625(17)$,  (including \emph{alsq} terms, see Figure~\ref{Fig:3} captain and Eqn.~\ref{eqn1} -- \ref{eqn5} for notation), 
and $f_0^K = 0.768(12)$ (including \emph{alsqe} terms), 
where errors shown are statistical only. 
To estimate the systematic error from chiral-continuum fits in a conservative manner, 
we take the difference between the central value of the result of the central fit and that of the fit (with $\chi^2/\text{dof.} \leq 1.0$) furthest from the central fit. 
These results are preliminary. 
Further investigations will finalize the study of the stability of fits under the inclusion of different NNLO terms and the analysis of systematic uncertainties.

\subsection{Error Budgets and Preliminary Results} 

Besides chiral-continuum extrapolations, 
other systematic errors include: 
\begin{description}
\item[ $\bullet$] One-loop partial quenching (PQ) effects on $f_0^K$, 
because of the partial-quenching of the strange quark mass. 
The one-loop chiral-logarithm term used is deduced from full QCD where $m_s^{val} = m_s^{sea}$, 
but for some of our data $m_s^{val} \ne m_s^{sea}$. 
To estimate this effect, 
$m_s$ in the formula is taken as $m_s^{val}$ or $m_s^{sea}$, 
and $f_0^K$ from the fits are compared. 
For our final results we plan to use partially-quenched $\chi$PT expressions. 
\item[ $\bullet$] Leading-order lattice discretizations at $O(\alpha_s^2 a^2)$, 
which are estimated on both the momentum and the mass of the light quark. 
The charm quark mass discretization effect is not considered at present, 
but will be estimated for the final result. 
\item[ $\bullet$] Nonequilibrated topological charge on the $a \approx 0.042$~fm ensemble~\cite{TopolgCh}. 
With such a small lattice spacing, 
the autocorrelation time of the topological charge is large and the topological sectors therefore not correctly sampled. 
The necessary corrections for both scalar form factors are deduced from SU(3) chiral perturbation theory, 
\begin{eqnarray} 
f_0 &=& f_0(\theta) - f_0'' (2\chi_T V)^{-1} ( 1 - \langle Q^2 \rangle (\chi_T V)^{-1} ) , \\
f_0'' &=& -1/4 ( m_l' m_s' / m_y)^2 (m_l' + 2 m_s')^{-2} , 
\end{eqnarray}
where $\theta$ is the vacuum angle, 
$\chi_T$ is the topological susceptibility, 
$Q$ is the topological charge operator, 
$V$ is the four-dimensional lattice volume, 
$m_{l,s}'$ are sea quark masses, 
and $m_y$ is the active light valence quark mass. 
\end{description}
\begin{table}[htbp]
\begin{center}
\begin{tabular}{ |c|c|c| }
\hline
Systematic errors	&	$f_0^{\pi}(0)$ (\%)	&	$f_0^K(0)$ (\%)	\\
\hline
chiral-continuum~fit stab.	&	2.15	&	1.47 \\
\hline
PQ effects	&	$N/A$	&	0.12 \\
\hline
Lattice scale	&	0.05	&	0.03 \\
\hline
Finite volume	&	0.04	&	<0.04 \\
\hline
Topology	&	0.005	&	<0.005 \\
\hline
\textbf{Total}	&	\textbf{2.15}	&	\textbf{1.48} \\
\hline
\end{tabular}
\end{center}
\caption{Error budgets (preliminary) of $f_0^{\{\pi, K\}}(0)$ in this work. 
Systematic errors are shown in percentage. 
The finite volume effect is estimated for $f_0^{\pi}(0)$ following the procedure in Ref.~\cite{FVQCD}, 
and its relative error is assumed to be smaller for $f_0^K(0)$. }
\label{Tab:3}
\end{table}

\begin{table}[htbp]
\begin{center}
\begin{tabular}{c c c}
\hline
\hline
	&	$f_+^\pi (q^2=0)$	&	$f_+^K (q^2=0)$ \\
\hline
This work	&	0.625(17)(13)	&	0.768(12)(11) \\	
\hline
FLAG	&	0.666(29)	&	0.747(19) \\
\hline
\hline
\end{tabular}
\end{center}
\caption{Results of $f_+^{\{\pi, K\}}(0)$, comparing between this work and the FLAG report~\cite{FLAG}. 
Results from this work are shown with both statistical (first) and systematic (second) errors. }
\label{Tab:4}
\end{table}

Table~\ref{Tab:3} contains preliminary error budgets of this work,  
in which errors from the lattice scale setting and finite-volume effects~\cite{FVQCD} are also included. 
The major systematic error is from chiral-continuum extrapolations. 
Our results of the vector form factor at zero-momentum transfer $f_+^{\{\pi, K\}}(0)$ are consistent with the FLAG report in 2016 within $2 \sigma$, 
as shown in Table~\ref{Tab:4}. 

Using the latest experimental results reported from the Heavy Flavour Averaging Group~\cite{HFAG}, 
the second row CKM matrix elements calculated from $f_+(0)$ in this work are: $|V_{cd}| = 0.2280(30)(78)$, $|V_{cs}| = 0.941(4)(20)$. 
Comparing between the test of second row CKM unitarity $|V_{cd}|^2 + |V_{cs}|^2 + |V_{cb}|^2$ from this work: 0.939(42), 
and that from averaged leptonic and semileptonic determinations~\cite{FLAG}: 1.04(3), 
one can see that the tension between the semileptonic determination from this work and the averaged result is around $2 \sigma$.

\section{Summary}
We calculate the $D$ meson semileptonic decay vector form factors at zero-momentum transfer, $f_+^{\{\pi, K\}}(q^2 = 0)$. 
Calculations are done on a set of HISQ ensembles generated by MILC collaboration, 
with lattice spacings ranging from 0.042~fm to 0.12~fm, 
as well as several degenerate up/down quark masses down to the physical value. 
Our preliminary results $f_+^\pi(0) = 0.625(17)(13)$ and $f_+^K(0) = 0.768(12)(11)$, 
where the first(second) error is statistical(systematic), 
are consistent with recent lattice-QCD determinations from the semileptonic decay process.

\section{Acknowledgements}
Computations for this work were carried out with resources provided by the USQCD Collaboration, the National Energy Research Scientific Computing Center, the Argonne Leadership Computing Facility, the Blue Waters sustained-petascale computing project, the National Institute for Computational Science, the National Center for Atmospheric Research, and the Texas Advanced Computing Center. USQCD resources are acquired and operated thanks to funding from the Office of Science of the U.S. Department of Energy. The National Energy Research Scientific Computing Center is a DOE Office of Science User Facility supported by the Office of Science of the U.S. Department of Energy under Contract No. DE-AC02-05CH11231. An award of computer time was provided by the Innovative and Novel Computational Impact on Theory and Experiment (INCITE) program. This research used resources of the Argonne Leadership Computing Facility, which is a DOE Office of Science User Facility supported under Contract DE-AC02-06CH11357. The Blue Waters sustained-petascale computing project is supported by the National Science Foundation (awards OCI-0725070 and ACI-1238993) and the State of Illinois. Blue Waters is a joint effort of the University of Illinois at Urbana-Champaign and its National Center for Supercomputing Applications. This work is also part of the ``Lattice QCD on Blue Waters'' and ``High Energy Physics on Blue Waters'' PRAC allocations supported by the National Science Foundation (award numbers 0832315 and 1615006) and used an allocation received under the ``Blue Waters for Illinois faculty'' program. This work used the Extreme Science and Engineering Discovery Environment (XSEDE), which is supported by National Science Foundation grant number ACI-1548562 [161]. Allocations under the Teragrid and XSEDE programs included resources at the National Institute for Computational Sciences (NICS) at the Oak Ridge National Laboratory Computer Center, The Texas Advanced Computing Center and the National Center for Atmospheric Research, all under NSF teragrid allocation TG-MCA93S002. Computer time at the National Center for Atmospheric Research was provided by NSF MRI Grant CNS-0421498, NSF MRI Grant CNS-0420873, NSF MRI Grant CNS-0420985, NSF sponsorship of the National Center for Atmospheric Research, the University of Colorado, and a grant from the IBM Shared University Research (SUR) program. 

A gift from Intel though the Intel Parallel Computing Center at Indiana University supported the work of R.L.. 
This work was supported in part by the U.S.\ Department of Energy under grants
No.\ DE-FG02-91ER40628 (C.W.B.),
No.\ DE-FC02-12ER41879 (C.D.),
No.\ DE-SC0010120 (S.G.),
No.\ DE-FG02-91ER40661 (S.G.),
No.\ DE-FG02-13ER42001 (A.X.K.),
No.\ DE-SC0015655 (A.X.K.),
No.\ DE-SC0010005 (E.T.N.),
No.\ DE-FG02-13ER41976 (D.T.);
by the U.S.\ National Science Foundation under grants
PHY14-14614 and PHY17-19626 (C.D.),
PHY14-17805 (J.L.), and
PHY13-16748 and PHY16-20625 (R.S.);
by the MINECO (Spain) under grants FPA2013-47836-C-1-P and FPA2016-78220-C3-3-P (E.G.);
by the Junta de Andaluc\'{\i}a (Spain) under grant No.\ FQM-101 (E.G.);
by the European Commission (EC) under Grant No.\ PCIG10-GA-2011-303781 (E.G.);
by the Fermilab Distinguished Scholars Program (A.X.K.);
by the U.K.\ Science and Technology Facilities Council (J.K.);
by the German Excellence Initiative and the European Union Seventh Framework Program under grant agreement No.~291763 as well as the
European Union Marie Curie COFUND program (J.K., A.S.K.).
Brookhaven National Laboratory is supported by the United States Department of Energy, Office of Science, Office of High Energy
Physics, under Contract No.\ DE-SC0012704.
Fermilab is operated by Fermi Research Alliance, LLC, under Contract No.\ DE-AC02-07CH11359 with the United States Department of
Energy, Office of Science, Office of High Energy Physics.

\end{document}